\documentclass[twocolumn,secnumarabic,amssymb, nobibnotes, aps, prc,superscriptaddress]{revtex4-1}

\usepackage{xcolor,soul}
\usepackage{bm}
\usepackage{graphicx}
\usepackage{amsmath}
\usepackage{epstopdf}
\usepackage{multirow}
\begin{document}

\title{Photoneutron cross sections for Ni isotopes: Toward understanding $(n,\gamma)$ cross sections relevant to the weak s-process nucleosynthesis}

\author{H.~Utsunomiya$^{\ast}$}
\affiliation{Department of Physics, Konan University, Okamoto 8-9-1, Higashinada, Kobe 658-8501, Japan}
\email[]{hiro@konan-u.ac.jp}

\author{T.~Renstr\o m}
\affiliation{Department of Physics, University of Oslo, N-0316 Oslo, Norway}

\author{G.~M~Tveten}
\affiliation{Department of Physics, University of Oslo, N-0316 Oslo, Norway}

\author{S.~Goriely}
\affiliation{Institut d'Astronomie et d'Astrophysique, Universit\'{e} Libre de Bruxelles, Campus de la Plaine, CP-226, 1050 Brussels, Belgium}

\author{S.~Katayama}
\affiliation{Department of Physics, Konan University, Okamoto 8-9-1, Higashinada, Kobe 658-8501, Japan}

\author{T.~Ari-izumi}
\affiliation{Department of Physics, Konan University, Okamoto 8-9-1, Higashinada, Kobe 658-8501, Japan}

\author{D.~Takenaka}
\affiliation{Department of Physics, Konan University, Okamoto 8-9-1, Higashinada, Kobe 658-8501, Japan}

\author{D.~Symochko}
\affiliation{Technische Universit\"at Darmstadt, Karolinenplatz 5, 64289 Darmstadt, Germany}

\author{B. V.~Kheswa}
\affiliation{Department of Physics, University of Oslo, N-0316 Oslo, Norway}
\affiliation{University of Johannesburg, Department of Applied physics and Engineering mathematics, Doornfontein, Johannesburg, 2028, South Africa}

\author{V.W. Ingeberg}
\affiliation{Department of Physics, University of Oslo, N-0316 Oslo, Norway}

\author{T.~Glodariu$^{\dagger}$}
\affiliation{Extreme Light Infrastructure Nuclear Physics, "Horia Hulubei" National Institute for Physics and Nuclear Engineering (IFIN-HH), 30 Reactorului, 077125 Bucharest-Magurele, Romania}
\thanks{Deceased 14 November 2017}

\author{Y.-W.~Lui}
\affiliation{Cyclotron Institute, Texas A\&M University, College Station, Texas 77843, USA}

\author{S.~Miyamoto}
\affiliation{Laboratory of Advanced Science and Technology for Industry, University of Hyogo, 3-1-2 Kouto, Kamigori, Ako-gun, Hyogo 678-1205, Japan}

\author{A.~C.~Larsen}
\affiliation{Department of Physics, University of Oslo, N-0316 Oslo, Norway}

\author{J.~E.~Midtb\o}
\affiliation{Department of Physics, University of Oslo, N-0316 Oslo, Norway}

\author{A.~G\"orgen}
\affiliation{Department of Physics, University of Oslo, N-0316 Oslo, Norway}

\author{S.~Siem}
\affiliation{Department of Physics, University of Oslo, N-0316 Oslo, Norway}

\author{L.~Crespo~Campo}
\affiliation{Department of Physics, University of Oslo, N-0316 Oslo, Norway}

\author{M.~Guttormsen}
\affiliation{Department of Physics, University of Oslo, N-0316 Oslo, Norway}

\author{S.~Hilaire}
\affiliation{CEA, DAM, DIF, F-91297 Arpajon, France}

\author{S.~P\'eru}
\affiliation{CEA, DAM, DIF, F-91297 Arpajon, France}

\author{A.~J.~Koning}
\affiliation{Nuclear Data Section, International Atomic Energy Agency, A-1400 Vienna, Austria}

\date{\today}%

\begin{abstract}
\vskip 0.5cm
Photoneutron cross sections were measured for $^{58}$Ni, $^{60}$Ni, $^{61}$Ni, and $^{64}$Ni at energies between the one-neutron and 
two-neutron thresholds using quasi-monochromatic $\gamma$-ray beams produced in laser Compton-scattering at the NewSUBARU synchrotron 
radiation facility.  The new photoneutron data are used to extract the $\gamma$-ray strength function above the neutron threshold complementing the information obtained by the Oslo method below the threshold. We discuss radiative neutron capture cross sections and the Maxwellian-averaged cross sections for Ni isotopes including $^{63}$Ni, a branching point nucleus along the weak s-process path. The cross sections are calculated with the experimentally constrained $\gamma$-ray strength functions from the Hartree-Fock-Bogolyubov plus quasi-particle-random phase approximation based on the Gogny D1M interaction for both $E1$ and $M1$ components and supplemented with the $M1$ upbend.

\end{abstract}

\maketitle

\section{Introduction}
\label{intro}
Nucleosynthesis of elements heavier than iron referred to as the slow neutron capture process, or s-process, is driven by repeated radiative neutron capture and $\beta$-decay. For the quest of understanding the s-process nucleosynthesis, radiative neutron capture cross sections are required as nuclear inputs to stellar models of such astrophysical sites, in particular the helium-shell burning of asymptotic giant branch stars for the main s-process component as well as the core helium-burning and carbon-burning phases in massive stars for the weak component \cite{Kapp11}. 
Although $(n,\gamma)$ data are rather well documented for stable nuclei \cite{Bao00}, those for radioactive nuclei at the s-process branching points remain an important research objective.

In the Hauser-Feshbach model of radiative neutron capture, the transmission of  a $\gamma$-ray of  energy $\epsilon_{\gamma}$ from a neutron-capture state at energy $E_x$ to state $\rho(E)$d$E$ is governed by the $\gamma$-ray strength function $f_{X\lambda}(\epsilon_{\gamma})$. Here $\rho(E)$ is the nuclear level density at $E$=$E_x$ - $\epsilon_{\gamma}$;  $X$ is either electric ($E$) or magnetic ($M$); and $\lambda$ is the radiation multipolarity.  In general, dipole radiation dominates over radiation of higher multipolarity for a given $\epsilon_{\gamma}$; so does electric over magnetic radiation for a given multipolarity.

The radiative neutron capture $(n,\gamma)$ and photoneutron $(\gamma,n)$ cross sections are interconnected by the $\gamma$-strength function ($\gamma$SF) through the Brink hypothesis \cite{Brink,Axel}. Here we refer to the equality of upward and downward electromagnetic transitions as the Brink hypothesis \cite{Brink,Axel,RIPL3}, apart from the hypothesis for photoabsorption from the ground state and an excited state in the standard Lorentzian model\cite{Brink}. The $(n,\gamma)$ reaction is governed by the downward $\gamma$SF $\overleftarrow{f_{X\lambda}}(\varepsilon_\gamma)$ at $\epsilon_{\gamma}$ $<$ $S_n$, the one-neutron separation energy, while the  $(\gamma,n)$ by the upward $\gamma$SF
$\overrightarrow{f_{X\lambda}}(\varepsilon_\gamma)$ at $\epsilon_{\gamma}$ $>$ $S_n$.  The Brink hypothesis assumes the approximate equality of $\overleftarrow{f_{X\lambda}}$  and $\overrightarrow{f_{X\lambda}}$ .

The $(\gamma,n)$ cross section vanishes at $S_n$ because the neutron transmission is inhibited at the threshold.  At energies excluding those immediately above $S_n$, the $(\gamma,n)$ cross section provides $\overleftarrow{f_{X\lambda}}(\varepsilon_\gamma)$ relevant to the radiative neutron capture with an absolute normalization.  This function of the $(\gamma,n)$ cross section is particularly useful for an absolute normalization of 
$\overleftarrow{f_{X\lambda}}(\varepsilon_\gamma)$  \cite{Toft10} deduced with the Oslo method \cite{Guttormsen1987, Guttormsen1996, Schi00, Larsen2011}, which is briefly summarized in Sec.IV.

The $\gamma$-ray strength function method \cite{Utsu10a} ($\gamma$SF method) has been devised to investigate systematically $(\gamma,n)$ and $(n,\gamma)$ cross sections over an isotopic chain.  Model $\gamma$-ray strength functions constrained by the $(\gamma,n)$ cross section are further tested against existing $(n,\gamma)$ cross sections.  This method effectively allows us to deduce the $(n,\gamma)$ cross section for radioactive nuclei involved in the isotopic chain as a result of the unified and comprehensive approach \cite{Utsu10b,Utsu11,Utsu13,Fili14,Nyhu15}. 

\begin{figure}
\begin{center}
\includegraphics[bb = 140 220 800 500, scale=0.45]{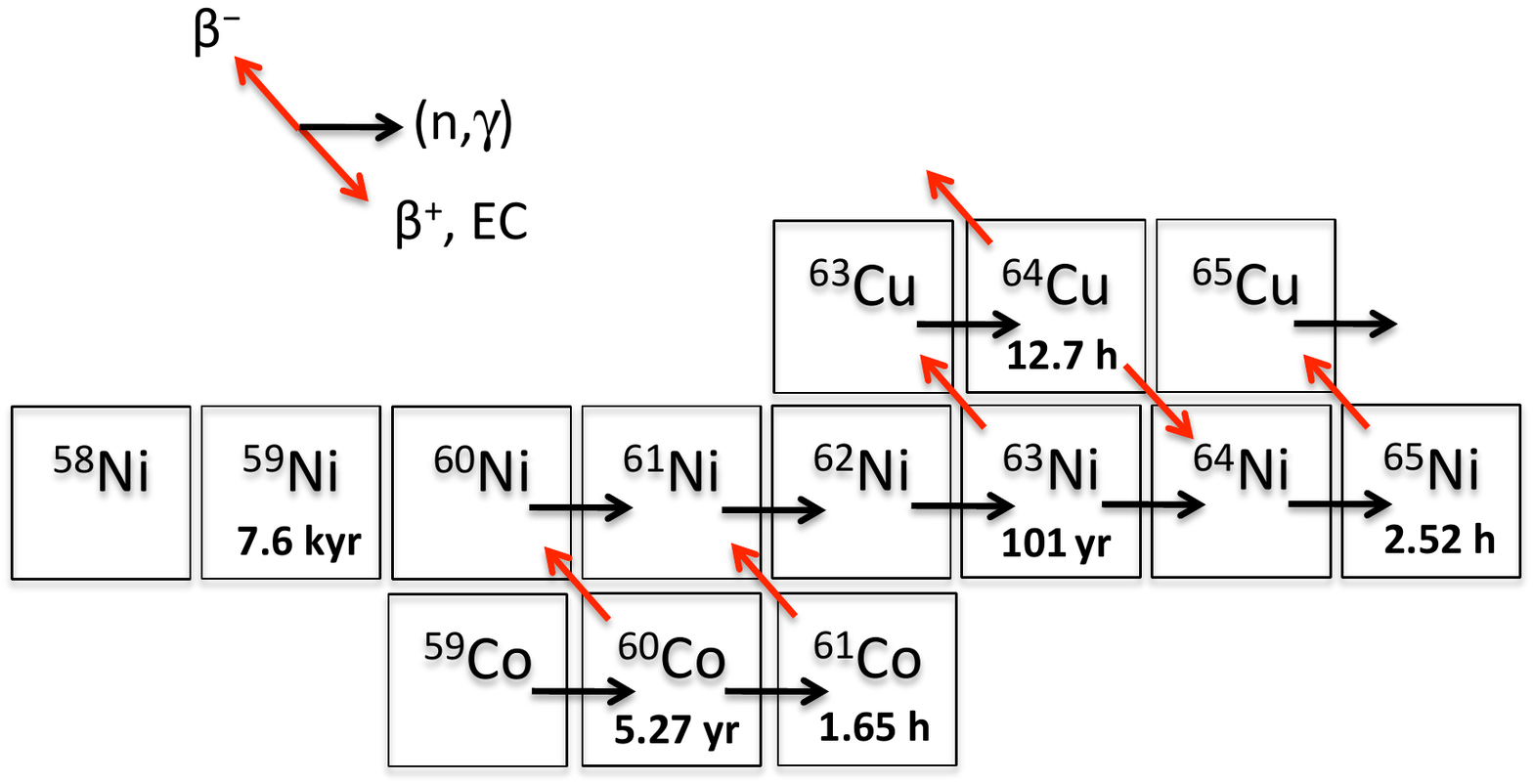}
\caption{(Color online) An excerpt of the chart of nuclei depicting Ni isotopes of the present research objectives in relation to the weak s-process path.  
}
\label{fig-Chart_nuclei}       

\end{center}
\end{figure}

In this paper, we report results obtained in the application of the $\gamma$-ray strength function method to the Ni isotopic chain, where 
 $(\gamma,n)$ cross sections for $^{58}$Ni, $^{60}$Ni, $^{61}$Ni, and $^{64}$Ni are combined with $\overleftarrow{f_{X\lambda}}(\varepsilon_\gamma)$  for $^{59}$Ni, $^{60}$Ni, $^{64}$Ni, and $^{65}$Ni deduced with the Oslo method.  Figure~\ref{fig-Chart_nuclei} shows an excerpt of the chart of nuclei depicting Ni isotopes of the present research objectives in relation to the weak s-process path.  We discuss $(n,\gamma)$ cross sections for $^{60}$Ni, $^{63}$Ni with the half-life ($T_{1/2}$) of 101 yr, and $^{64}$Ni of direct relevance to the weak s-process nucleosynthesis.  We also discuss $(n,\gamma)$ cross sections for $^{59}$Ni ($T_{1/2}$=7.6 $\times$ 10$^4$ yr) in the context of the $\gamma$SF method.  

\section{Experimental procedure}
\label{sec-ExpProc}

Quasi-monochromatic $\gamma$-ray beams were produced by laser-Compton scattering (LCS) at the NewSUBARU synchrotron radiation facility. Electrons at 974~MeV were injected from the electron linear accelerator into the NewSUBARU storage ring and stored with up to 300~mA at 500~MHz in the top-up operation. The INAZUMA (Nd:YVO$_4$, 1064~nm) laser was operated in the Q-switch mode at the frequency 20 kHz. After the injection, the electron beam energy was decelerated, in steps, from 974~MeV down to 663~MeV to produce LCS $\gamma$-ray beams in the range of 16.93 - 8.00~MeV. The beam was accelerated, in steps, from 974~MeV up to 1113~MeV to produce LCS $\gamma$-ray beams in the range of 16.93 - 22.02~MeV. 
The quoted electron energy is the nominal energy which has been calibrated with the accuracy of the order of 10$^{-5}$ \cite{Utsu14}. The quoted $\gamma$ energy is the maximum energy of the quasi-monochromatic $\gamma$-ray beam.  

An external transistor-transistor logic (TTL) gate with 80~ms pulse width was applied to switch on the INAZUMA laser at 10~Hz frequency, producing a macro time-structure of beam-on for 80~ms followed by beam-off for 20~ms.  
The LCS $\gamma$-ray beam produced in collisions with electron bunches at 500~MHz has the same micro (20~kHz) and macro (80~ms/20~ms) time-structure as the laser.   

Enriched samples of $^{58}$Ni, $^{60}$Ni, $^{61}$Ni, and $^{64}$Ni were shaped into disks with diameters 7.8 - 10.0~mm and mounted in windowless cylindrical holders made of aluminum. The Ni sample was placed at the center of the High-Efficiency Neutron Detector consisting of three concentric rings of 4, 8, and 8 $^3$He-filled proportional counters embedded in a polyethylene moderator \cite{Fili14} and irradiated with the LCS $\gamma$-ray beam at energies between the one- and two-neutron thresholds. The Ni samples used are listed in Table I along with the neutron threshold energies.    
Reaction plus background neutrons were measured for every 80 ms of LCS $\gamma$ beam on, while background neutrons were measured during every 20 ms of beam off.  The detection efficiency of the triple-ring neutron detector was recently remeasured using a calibrated $^{252}$Cf source with an emission rate of 2.27 
$\times$ 10$^4$ s$^{-1}$ with 2.2 \% uncertainty at the National Meteorology Institute of Japan \cite{Nyhu15}.  Details of the neutron detection can be found in Ref. \cite{Fili14}. 

For the $\gamma$-ray flux determination, the pulsed LCS $\gamma$-ray beam was measured with a large-volume (8" $\times$ 12") NaI(Tl) detector. The number of $\gamma$-rays, $N_{\gamma}$, was determined from multi-photon spectra based on the Poisson-fitting method \cite{Kii99,Toyo00} or the so-called pile-up method \cite{Kond11}.  Recently, the accuracy of the experimental formula \cite{Kond11} used for the $\gamma$-flux determination was thoroughly investigated with the Poisson-fitting method, showing that the inherent uncertainty of the method is less than 0.1 \% \cite{Utsu18} provided that multi-photon spectra are free from quenching at the photomultiplier tube of the NaI(Tl) detector .   

The energy distribution of the LCS $\gamma$-ray beam was determined by best reproducing response functions of a 3.5" $\times$ 4.0" LaBr$_3$(Ce) detector to LCS $\gamma$-ray beams with a GEANT4 code \cite{geant4ref} that incorporated the kinematics of collisions between laser photons and electrons and $\gamma$-ray beam-transport through collimators to the detector. 

The incident energy distribution was used to determine the cross section, $\sigma(\varepsilon_{\gamma})$, as a function of incident photon energy, $\varepsilon_{\gamma}$. More specifically, the incoming photon-beam spectra were used to determine $D^{E_{\rm max}}$, the energy distribution of
the beam normalized to unity, $\int_{S_n}^{E_{\rm max}} D^{E_{\rm max}}d\varepsilon_{\gamma}= 1$. The measured $\sigma^{E_{\rm max}}_{\rm exp}$ for each electron beam energy can be expressed as
\begin{equation}
\sigma^{E_{\rm max}}_{\rm exp}=\int_{S_n}^{E_{\rm max}}D^{E_{\rm max}}(\varepsilon_{\gamma})\sigma(\varepsilon_{\gamma})d\varepsilon_{\gamma}=\frac{N_n}{N_tN_{\gamma}\xi\epsilon_n g}.
\label{eq:cross1}
\end{equation}
Here,  $N_t$ gives the number of target nuclei per unit area, $\epsilon_n$ represents the neutron detection efficiency, and $\xi=(1-e^{-\mu t})/(\mu t)$ gives a correction factor for the self-attenuation effect in a thick-target measurement where $\mu$ is the linear attenuation coefficient.  Finally, the factor $g$ represents the fraction of the $\gamma$ flux above $S_n$. 
\begin{table}
\caption{Nickel targets used in the present measurement. }
\begin{center}
  \begin{tabular}{ l c c c c c }
    \hline
     \hline
    isotope & abundance & enrichment  & thickness  & $S_n$ & $S_{2n}$ \\ 
     & [$\%$] & [$\%$] & [g/cm$^2$] & [MeV] & [MeV] \\ \hline 
    $^{58}$Ni & 68.08 & 99.80 &  1.550  & 12.22 & 22.47\\
    $^{60}$Ni  & 26.22 & 99.5 $\pm$ 0.1 & 1.041 & 11.39 & 20.39\\
    $^{61}$Ni  & 1.14 & 91.14 $\pm$ 0.05  & 0.608 & 7.82 & 19.21\\
    $^{64}$Ni  & 0.93 & 95.4 $\pm$ 0.3  & 1.023 & 9.66 & 16.50\\
    \hline
     \hline
  \end{tabular}
\end{center}
\end{table}


\section{Unfolding photoneutron cross sections}
\label{subsec-DataRed}
The measured $\sigma^{E_{\rm max}}_{\rm exp}$ is, as shown in Eq.(\ref{eq:cross1}), convoluted by the energy distribution of the beam. We want to extract the deconvoluted, $\varepsilon_{\gamma}$ dependent, photo-neutron cross section, $\sigma(\varepsilon_{\gamma})$, from the integral of Eq.~(\ref{eq:cross1}). We approximated the integral in Eq.~(\ref{eq:cross1}) with a sum for each $\gamma$-beam profile, thus enabling us to express the problem as a set of linear equations
\begin{equation}
\sigma_{\rm f }=\bf{D}\sigma,
\end{equation}
where $\sigma_{\rm f}$ is the cross section folded with the beam profile {\bf D}.  
The indexes $i$ and $j$ of the matrix element $D_{i,j}$ corresponds to $E_{\rm max}$ and $\varepsilon_{\gamma}$, respectively.
The set of equations is given by
\begin{equation}
\begin{pmatrix}\sigma_{\rm{1}}\\\sigma_{\rm{2}}\\ \vdots \\ \sigma_N \end{pmatrix}_{\rm f}\\\mbox{}=
\begin{pmatrix}D_{ 11} & D_{ 12}& \cdots &\cdots &D_{ 1M} \\ D_{ 21} & D_{ 22}&
\cdots & \cdots &D_{ 2M} \\ \vdots &\vdots & \vdots & \vdots & \vdots \\ D_{ N1} & D_{ N2}& \cdots & \cdots &D_{ NM}\end{pmatrix}
\begin{pmatrix}\sigma_{1}\\\sigma_{2}\\ \vdots \\ \vdots \\\sigma_{M} \end{pmatrix},
\label{eq:matrise_unfolding}
\end{equation}
where each row of $\bf{D}$ corresponds to the simulated $\gamma$
beam profile corresponding to $E_{\rm max}$. 

As the system of linear equations in Eq.~(\ref{eq:matrise_unfolding}) is under-determined, the $\sigma$ vector cannot be found by matrix inversion. We determine $\sigma$ through an iterative folding method that can be summarized as follows:

\begin{itemize}

\item [1)] As our starting point, we choose for the 0th iteration, a constant trial function $\sigma^0$.
This initial vector
is multiplied with $\bf{D}$, and we get the 0th folded vector $\sigma^0_{\rm f}= {\bf D} \sigma^{0}$.

\item[2)] The next trial input function, $\sigma^1$, is established by adding the difference of
the experimentally measured spectrum, $\sigma_{\rm{exp}}$, and the folded spectrum, $\sigma^0 _{\rm f}$,
to $\sigma^0$. In order to be able to add the folded and the input vector together, we first perform a spline
interpolation on the folded vector, then interpolate so that the two vectors have equal dimensions. Our new input vector is:

\begin{equation}
\sigma^1 = \sigma^0 + (\sigma_{\rm{exp}}-\sigma^0 _{\rm f}).
\end{equation} 

\item[3)] The steps 1) and 2) are iterated $i$ times giving
\begin{eqnarray}
\sigma^i_{\rm f} &=& {\bf D} \sigma^{i}
\\
\sigma^{i+1}     &=& \sigma^i + (\sigma_{\rm{exp}}-\sigma^i _{\rm f})
\end{eqnarray}
until convergence is achieved. This means that
$\sigma^{i+1}_{\rm f} \approx \sigma_{\rm exp}$ within the statistical errors.
In order to quantitatively check convergence, we calculate the reduced $\chi^2$ of $\sigma^{i+1}_{\rm f}$ and
$\sigma_{\rm{exp}}$ after each iteration.
\end{itemize}

\begin{figure}
\begin{center}
\includegraphics[scale=0.60]{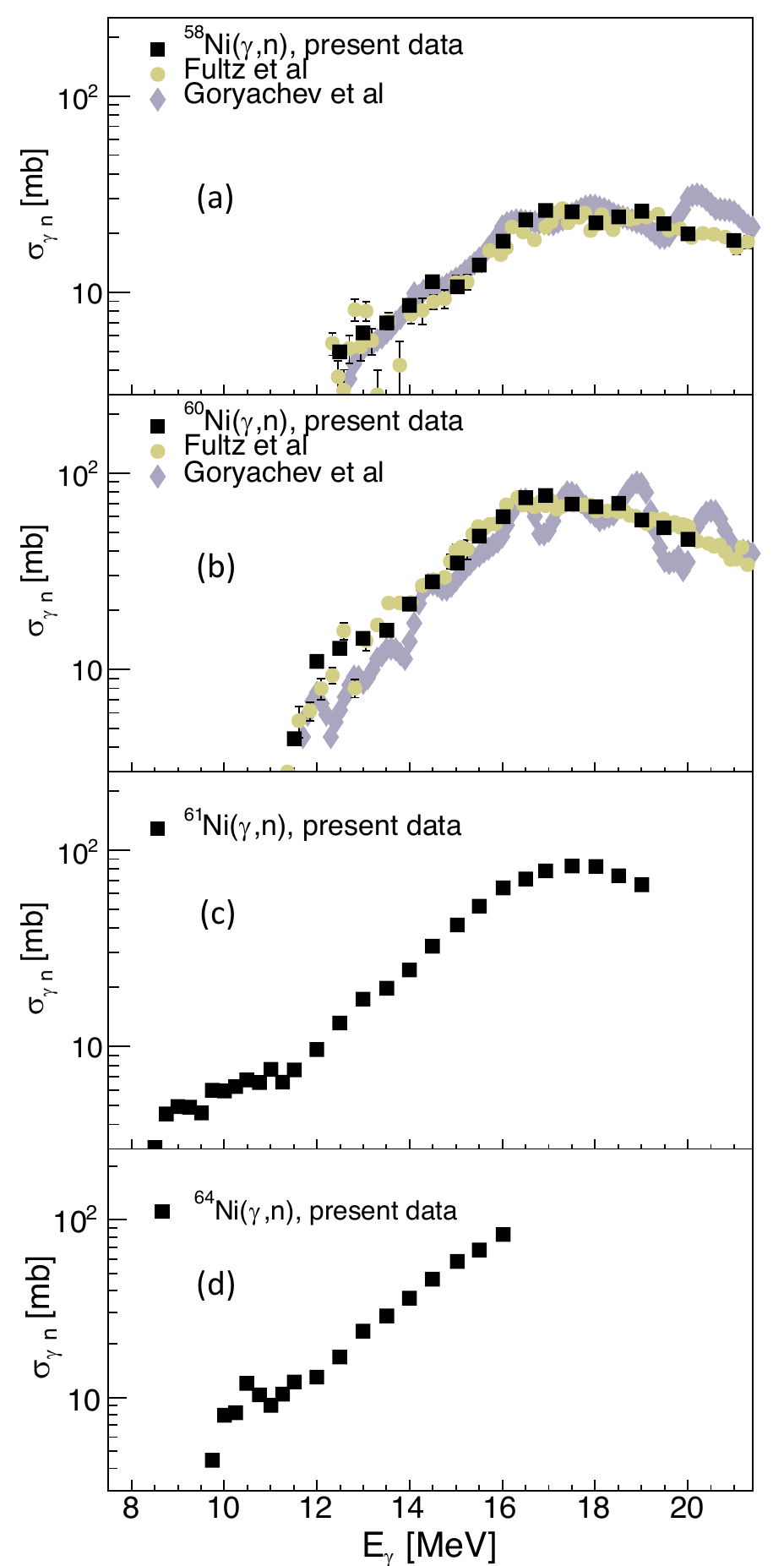}  
\caption{(Color online) (a,b) the $(\gamma, n)$ cross sections for $^{58,60}$Ni in comparison with the preceding data \cite{Goryachev68,Fultz74}. (c,d) the $(\gamma, n)$ cross sections for $^{61,64}$Ni measured for the first time. The error bars include both systematic and statistical uncertainties.}
\label{fig_all_ni}
\end{center}
\end{figure}

We stopped when the reduced $\chi^2$ was close to unity. Five iterations were found to be sufficient for the four data sets included in this work. To avoid spurious fluctuations, we applied a segmented sliding-average smoothing where the smoothing width was varied from 200 keV for the lowest photon energies and 400 keV for the highest. 

In order to give an estimate of the uncertainty in the unfolded cross sections, we have defined upper/lower limits of the monochromatic cross sections by adding/subtracting the errors to the measured cross section values. The upper and lower limits are unfolded separately to obtain error bars that take into account the propagation of errors through the folding method. 
In the error propagation, uncertainties of the neutron detection efficiency, (2.2\%), and $\gamma$-flux, (1 \%), were propagated as well as the neutron counting statistics.  Note that we took into account 1 \% uncertainty for the $\gamma$-flux associated with subtraction of background $\gamma$-rays though the inherent uncertainty of the Poisson-fitting method \cite{Utsu18} is negligible (less than 0.1 \%).  Thus, the systematic uncertainty in the absolute efficiency calibration of the neutron detector is included in the error bar of the $(\gamma,n)$ cross section.  

Fig.~\ref{fig_all_ni} shows the deconvoluted cross sections of $^{58,60,61,64}$Ni in comparison with the existing photoneutron data on $^{58}$Ni and $^{60}$Ni \cite{Goryachev68,Fultz74}. 
We provide the first data of $(\gamma,n)$ cross sections for two rare isotopes, $^{61}$Ni and $^{64}$Ni. 
The $^{58}$Ni and $^{60}$Ni data with improved accuracy near neutron threshold are rather close to the data of Fultz et al. \cite{Fultz74}, while especially the $^{60}$Ni data show discrepancies from the bremsstrahlung data \cite{Goryachev68}. The present measurement has confirmed structures around the top of the giant dipole resonance (GDR) previously seen in $^{58}$Ni and $^{60}$Ni \cite{Fultz74}.  We remark that the $(\gamma,p)$ channel contributes substantially to the decay of the GDR in $^{58}$Ni, resulting in a much reduced $(\gamma,n)$ cross section for this isotope. 


\section{$\gamma$-ray strength below $S_n$}
The isotopes $^{59,60,64,65}$Ni have been studied in charged particle experiments at the Oslo Cyclotron Laboratory (OCL). 
The experimental setup at OCL consisted of the SiRi Particle-Telescope System for light-ion induced nuclear reactions and the NaI(Tl) scintillator array CACTUS for $\gamma$ ray detection \cite{siri}. SiRi has eight trapezoidal modules that are mounted at 5 cm distance from the target, covering 8 forward angles. The thin front $\delta E$ detectors (130 micrometer) are segmented into eight pads, determining the reaction angle for the outgoing charged ejectile. The reaction ejectiles were identified by the $\delta E-E$ technique, thanks to the thick back E detectors (1550 micrometer). In coincidence with the ejectiles that provide information on the excitation energy, $\gamma$ rays were measured with CACTUS totaling 28 5"$\times$5" collimated NaI(Tl) scintillator detectors. Four reactions were used; $^{60}$Ni($^3$He,$\alpha\gamma$) and $^{60}$Ni($^3$He,$^3$He'$\gamma$) with 38 MeV $^3$He beam,  $^{64}$Ni(p,p'$\gamma$) with 16 MeV protons and finally$^{64}$Ni(d,p'$\gamma$) with 12.5 MeV deuterons. The targets were self-supporting enriched foils with thickness $\approx$ 2.0 mg/cm$^2$.

The $\gamma$-ray strength function and nuclear level density of these isotopes were extracted simultaneously from particle-$\gamma$ coincidence data through the application of the Oslo method \cite{Guttormsen1987, Guttormsen1996, Schi00, Larsen2011}. These $\gamma$-ray strength function results are combined with the $\gamma$-ray strength function extracted above $S_n$ from the ($\gamma$,n) cross sections. The Oslo method is based upon an iterative technique for extracting the shape of the primary $\gamma$-ray spectra from unfolded excitation energy (correcting for the detector response), $E_x$-tagged $\gamma$-ray spectra.  The primary $\gamma$-rays are the first $\gamma$s emitted in cascades and this set of $E_x$-tagged spectra result in a so-called first generation matrix, $P(E_x,\varepsilon_{\gamma})$.
In the analysis we assume that all available spins are populated and that $\gamma$-ray decay can be assumed to be of dipole character. 

The $P(E_x,\varepsilon_{\gamma})$ matrix can be viewed as a decay probability matrix and by making use of the assumption that the $\gamma$ ray transmission coefficient only depends on $\varepsilon_{\gamma}$,  in keeping with the generalized Brink hypothesis \cite{Brink,GuttormsenPRL2016}, the following decomposition can be carried out by a least-square fitting procedure
\begin{equation} \label{eq:rhotausep}
P(\varepsilon_{\gamma},E_x) \propto \rho(E_x - \varepsilon_{\gamma}) \mathcal{T}(\varepsilon_{\gamma}),
\end{equation}
where $ \rho(E_x - \varepsilon_{\gamma})$ is the nuclear level density at the excitation energy of the nucleus after a $\gamma$-ray with energy $\varepsilon_{\gamma}$ has been emitted and $\mathcal{T}(\varepsilon_{\gamma})$ is the $\gamma$-ray transmission coefficient \cite{Schi00}. Only the excitation energy range that correspond to statistical decay is included in the analysis. 

Once one solution for $\rho(E_x)$ and $\mathcal{T} (\varepsilon_{\gamma})$ was obtained, it has been shown in Ref. \cite{Schi00} that this solution can yield infinitely many by the following transformations
\begin{equation}\label{eq:norm1}
\tilde{\rho}(E_{x}-\varepsilon_{\gamma}) = {\cal A} \exp[\alpha(E_{x}-\varepsilon_{\gamma})]\rho(E_{x}-\varepsilon_{\gamma}), 
\end{equation}
\begin{equation}\label{eq:norm2}
\tilde{\mathcal{T}} (\varepsilon_{\gamma}) = {\cal B}\exp(\alpha \varepsilon_{\gamma})\mathcal{T} (\varepsilon_{\gamma}). 
\end{equation}
This means that the functional form of $ \rho(E_x - \varepsilon_{\gamma})$ and $\mathcal{T}(\varepsilon_{\gamma})$ are determined in the above mentioned procedure. The common slope and the absolute scales of the nuclear level density and $\gamma$-transmission coefficient, respectively, are found by normalizing to auxiliary data. The level density, $\tilde{\rho}(E_x - \varepsilon_{\gamma})$, is normalized by a least square fit of the parameters $\alpha$ and ${\cal A}$ to the level density found by counting discrete levels up to the excitation energy where the level scheme is considered complete and a data point at $S_n$ given by the level density at $S_n$, $\rho(S_n)$. The latter is found by calculating $\rho(S_n)$ from the average level spacing, $D$, using a model for the spin distribution of states. For $^{59}$Ni, $^{60}$Ni and $^{65}$Ni, experimental $D_0$ values from s-wave neutron resonances are available
from the RIPL-3 database \cite{RIPL3}.
In the case of $^{64}$Ni, $\rho(S_n)$ was estimated from the systematics of the Ni-isotopes and the normalization of the $\gamma$-ray strength function was adjusted to be in agreement with models for the GDR-tail. To calculate the level density at the neutron separation energy, $\rho(S_n)$, from $D$, we need to make use of a model describing the total spin distribution at $S_n$. The standard spin distribution of  Refs. \cite{spin1,spin2} was adopted with the spin cutoff values, 
$\sigma(S_n)$, provided in Table II together with the other values used for the normalization.

\begin{table}[]
\caption{The experimental information (the spin cutoff parameter, level density, and average radiative width)  used for the normalization of the $\gamma$-ray strength function. }
\begin{tabular}{lcccc}
    \hline
isotope  	 & $S_n$ {[}MeV{]} & $\sigma(S_n)$ & $\rho(S_n)$ MeV$^{-1}$ & $\langle\Gamma_{\gamma,0}\rangle$  {[}meV{]}\\
     \hline
$^{59}$Ni &        6.128             &     4.0                &     2536(520)            &    2030(800)         \\
$^{60}$Ni &        8.598            &     4.3                &     5349(2031)           &  2200(700)          \\
$^{64}$Ni  &       9.657            &     3.5                &     2620                      &       -   \\     
                   &                              &     4.1                 &     3470                       &      -    \\     
$^{65}$Ni  &       6.098            &     3.5                &      1120(110)        &  1090 $\pm$ 550 \\
    \hline          
\end{tabular}
\end{table}

The transmission coefficient is normalized according to 
\begin{equation}
\langle\Gamma_{\gamma}\rangle=\frac{1}{2\pi\rho(S_n)} \sum_{I_f} \int_{0}^{S_n} B\mathcal{T}(\varepsilon_{\gamma})\rho(S_n-\varepsilon_{\gamma},I_f)d\varepsilon_{\gamma},
\end{equation}
where $I_f$ is the spin of the final state, $\langle\Gamma_{\gamma}\rangle$ is the average radiative width and $\rho(S_n-\varepsilon_{\gamma},I_f)$ is the normalized level density weighted by the spin distribution. Again, experimental values of $\langle\Gamma_{\gamma}\rangle$ are available for $^{59}$Ni, $^{60}$Ni and $^{65}$Ni and the $\langle\Gamma_{\gamma}\rangle$ of $^{64}$Ni was estimated from systematics. Finally, the $\gamma$-ray strength function is determined by
\begin{equation}
f(\varepsilon_{\gamma}) = \frac{\mathcal{T}(\varepsilon_{\gamma})}{2\pi \varepsilon_{\gamma}^3},
\end{equation}
making use of the dominance of dipole radiation in the considered $E_x$ region~\cite{Kopecky2017,Larsen2013}. Further details on the experiments and analysis are provided in Refs. \cite{Crespo2016,Crespo2017,Renstrom2018}.  

\section{Theoretical interpretation}
\label{sec-results}

\begin{figure}
\begin{center}
\includegraphics[scale=0.30]{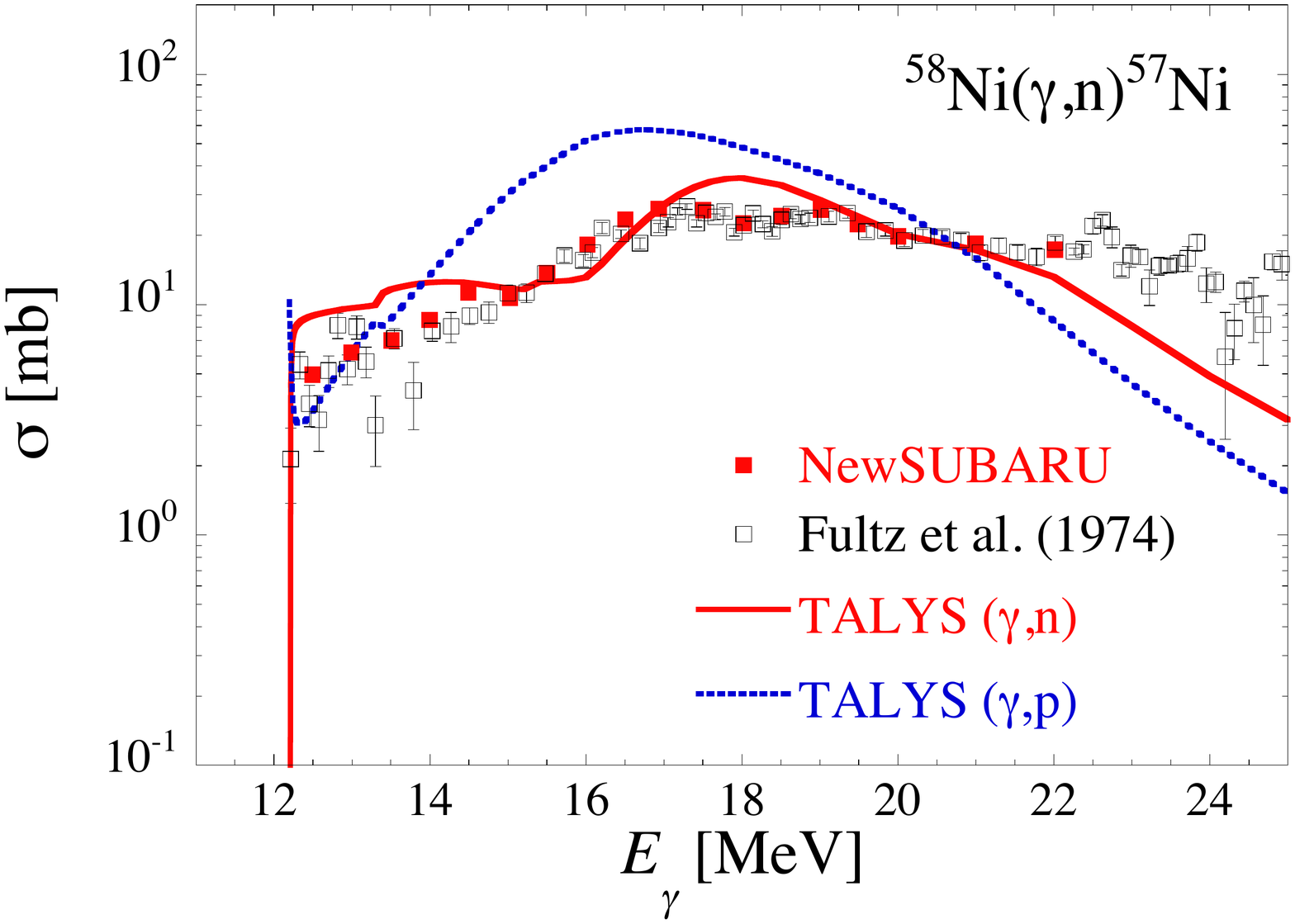}  
\caption{(Color online) Comparison between the experimental $^{58}$Ni$(\gamma,n)^{57}$Ni cross section (red squares) and the TALYS predictions for both the $(\gamma,n)$ (solid red line) and $(\gamma,p)$ (dotted blue line) cross sections. Experimental data from Ref.~\cite{Fultz74} (black open squares) are also included.}
\label{fig_58Nign}
\end{center}
\end{figure}

\begin{figure}
\begin{center}
\includegraphics[scale=0.40]{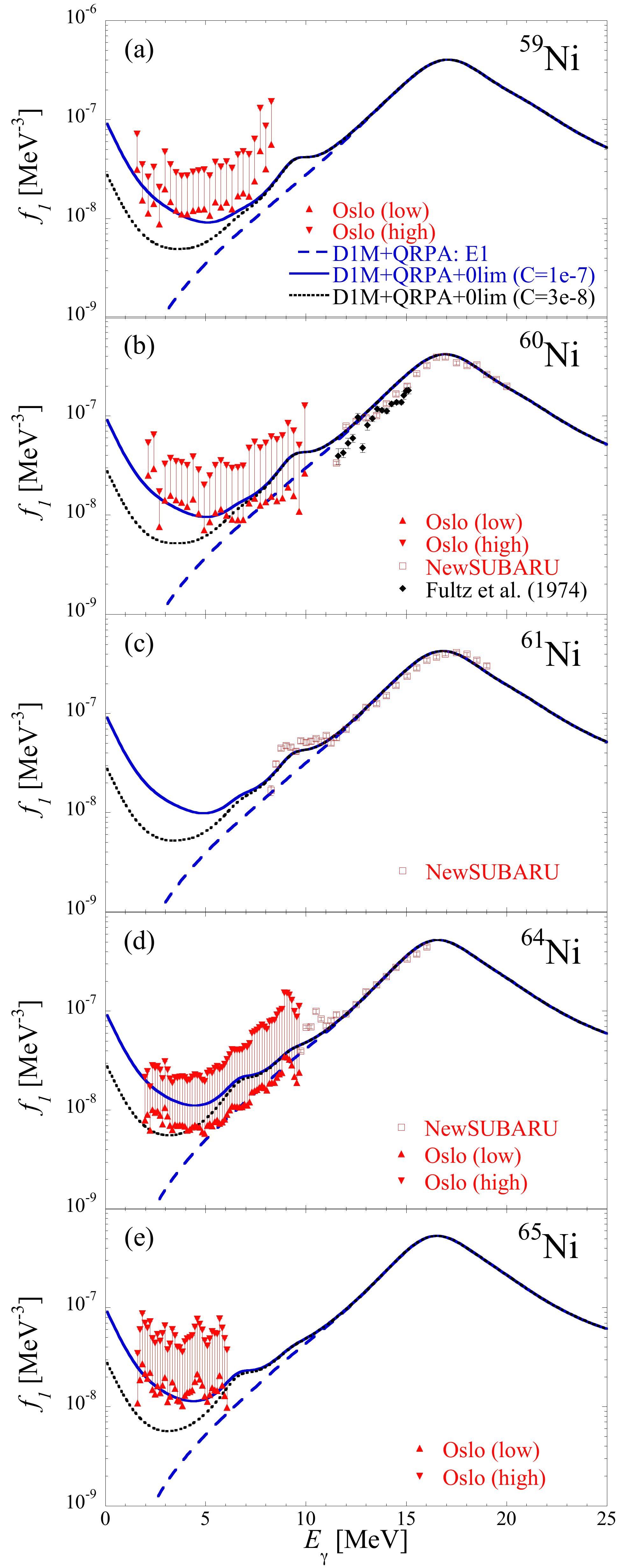}  
\caption{(Color online) (a-e) $\gamma$SF for the $^{59,60,61,64,65}$Ni isotopes. The red triangles correspond to the upper and lower limits of the $\gamma$SF extracted from the present Oslo data and the red open squares to the present NewSUBARU photoneutron data. The dashed blue curve represents the D1M+QRPA $E1$ strength and the black dotted (blue full) line the D1M+QRPA+0lim $E1$+$M1$ dipole strength obtained with $C=3\cdot 10^{-8}{\rm MeV^{-3}}$ ($C=10^{-7}{\rm MeV^{-3}}$). The $\gamma$SF of $^{64,65}$Ni are taken from \cite{Crespo2016,Crespo2017} (red triangles). The $\gamma$SF extracted from the $^{60}$Ni($\gamma$,n) data of Fultz et al. \cite{Fultz74} (black diamonds) is also shown in panel (b).}
\label{fig_gsf}
\end{center}
\end{figure}

The present photoneutron and Oslo data are now used to test some specific nuclear models commonly applied for systematic calculations in nuclear astrophysics applications, namely the mean field plus quasiparticle random phase approximation (QRPA) calculations  as well as the statistical Hauser-Feshbach model of nuclear reactions. The $\gamma$SF deduced from the Oslo data is complemented by the one extracted from the photoneutron cross section, $\sigma(\varepsilon_{\gamma})$, through the expression
\begin{equation}
f(\varepsilon_{\gamma})=\frac{1}{3\pi^2\hbar^2c^2}\frac{\sigma(\varepsilon_{\gamma})}{\varepsilon_{\gamma}}~.
\label{eq_gn}
\end{equation}
Note, however, that, in the vicinity of the neutron threshold, such expression does not hold because of the competition of the weak neutron channel. Similarly, if the photoproton emission or other  channels like $(\gamma,2n)$ compete with the $(\gamma,n)$ channel, the strength function cannot be extracted from Eq.~(\ref{eq_gn}) and in these cases, the Hauser-Feshbach formalism needs to be applied. This is in particular the case for $^{58}$Ni$(\gamma,n)^{57}$Ni, for which the $(\gamma,p)$ channel dominates the photoemission up to $\sim 21$~MeV, as shown in Fig.~\ref{fig_58Nign}. In this case, the cross section is highly sensitive to the adopted neutron and proton optical potential (here we used the standard Koning \& Delaroche potential \cite{Koning03}) and the extraction of the $\gamma$SF cannot be easily performed. Note that the TALYS code is considered in this paper for estimating the photoneutron and radiative neutron capture cross sections and reaction rates of astrophysical interest \cite{Koning12}. As detailed below, we will consequently concentrate on the $\gamma$SF of the $^{59,60,61,64,65}$Ni isotopes and on the available radiative neutron capture cross sections in the keV region, providing information on the relevance of nuclear properties such as the $\gamma$SF on cross sections of astrophysical interest.

\begin{figure}
\begin{center}
\includegraphics[scale=0.4]{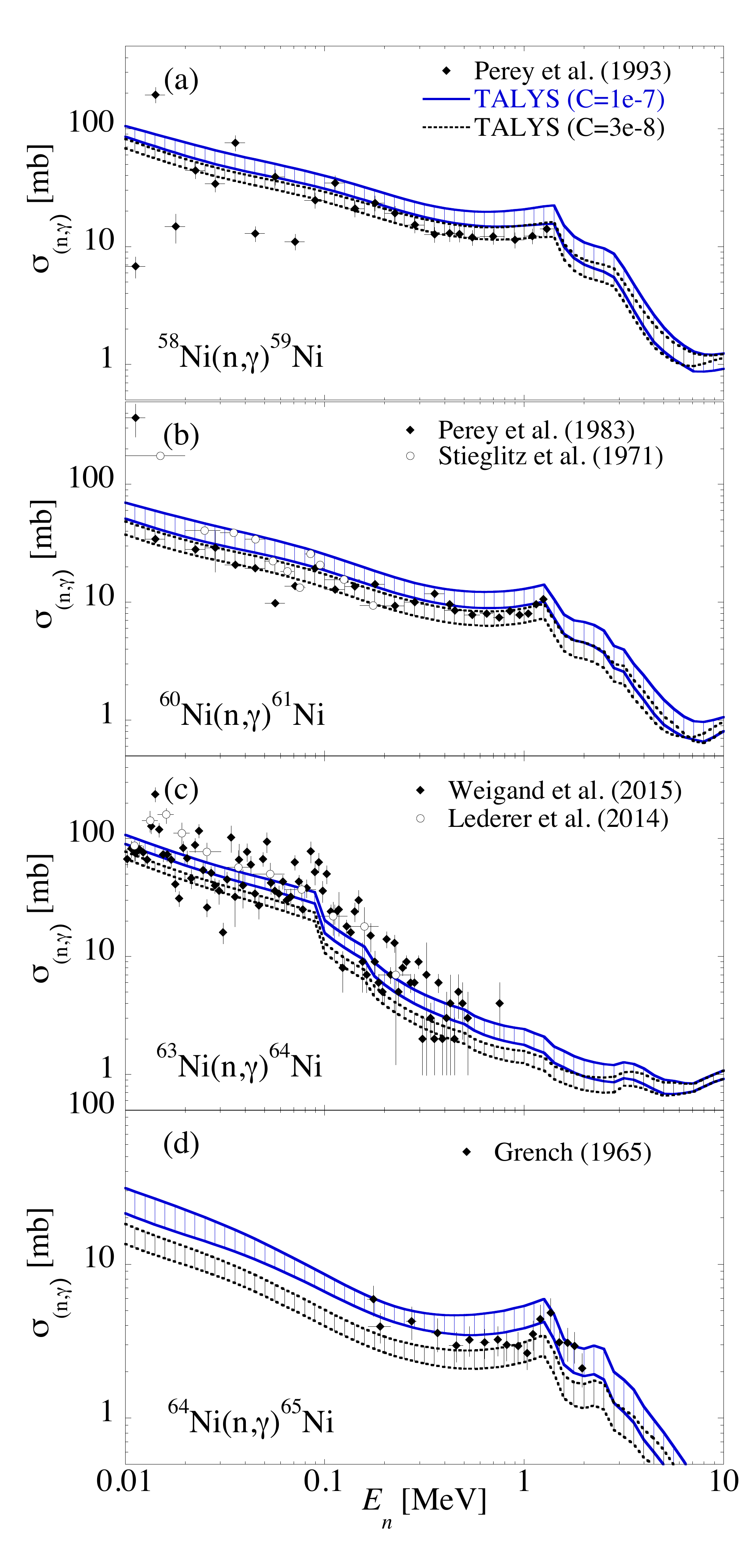}  
\caption{(Color online) radiative neutron capture cross section for the (a) $^{58}$Ni, (b) $^{60}$Ni, (c) $^{63}$Ni, and (d) $^{64}$Ni. The full (dotted) line corresponds to the TALYS calculation obtained with the D1M+QRPA+0lim dipole strength obtained with $C=3\cdot  10^{-8}{\rm MeV^{-3}}$ ($C=10^{-7}{\rm MeV^{-3}}$). Experimental data are taken from \cite{Perey93,Perey83,Stieglitz71,Weigand15,Lederer14,Grench65}. The hashed areas correspond to the prediction uncertainties associated with different nuclear level density models. }
\label{fig_ng}
\end{center}
\end{figure}

\begin{figure}
\begin{center}
\includegraphics[scale=0.4]{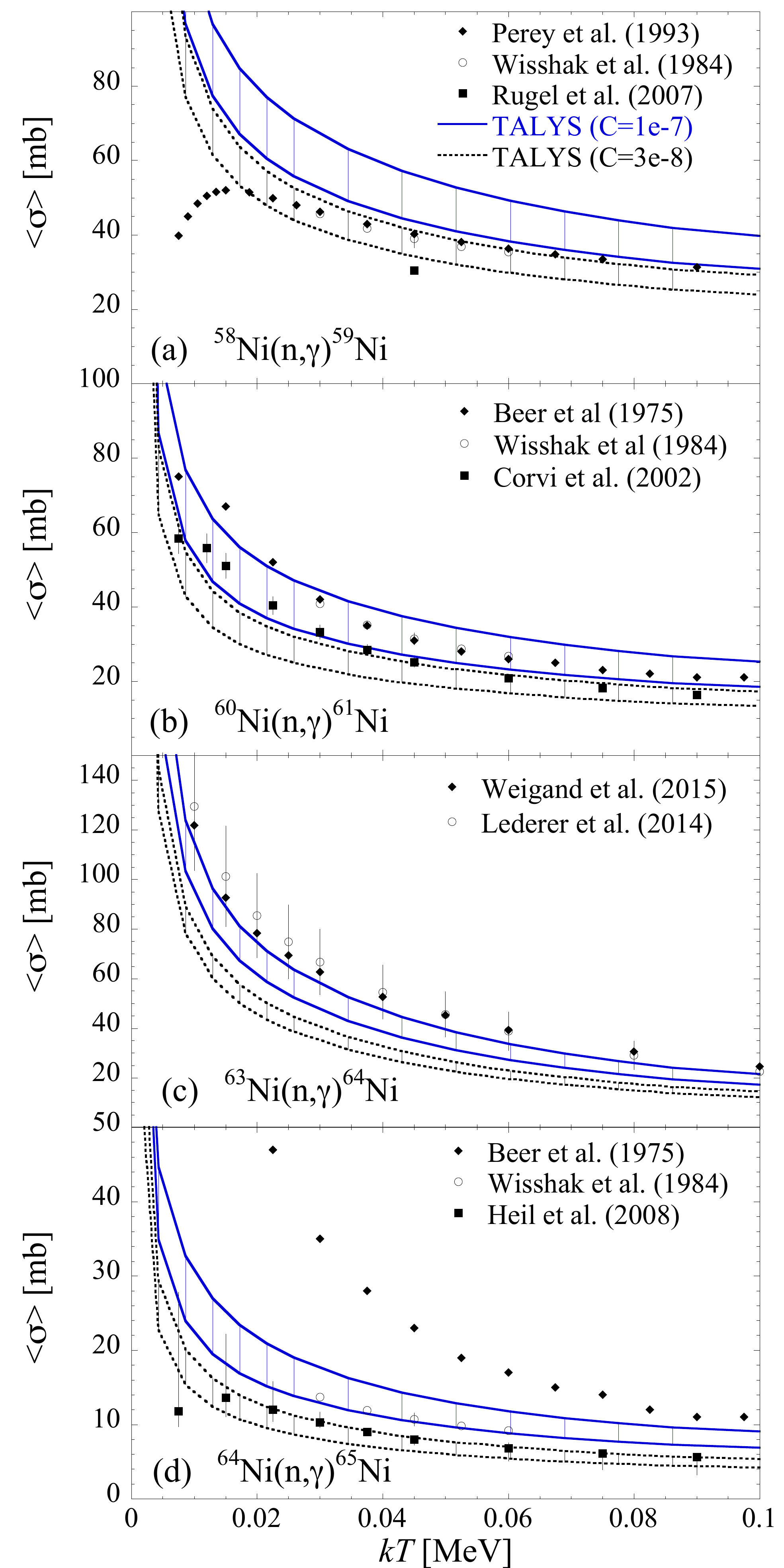}  
\caption{(Color online)  Same as Fig.~\ref{fig_ng} for the radiative neutron MACS. Experimental data are taken from \cite{Perey93,Wisshak84,Rugel07,Corvi02,Heil08,Beer75}.}
\label{fig_macs}
\end{center}
\end{figure}

\begin{figure}
\begin{center}
\includegraphics[scale=0.4]{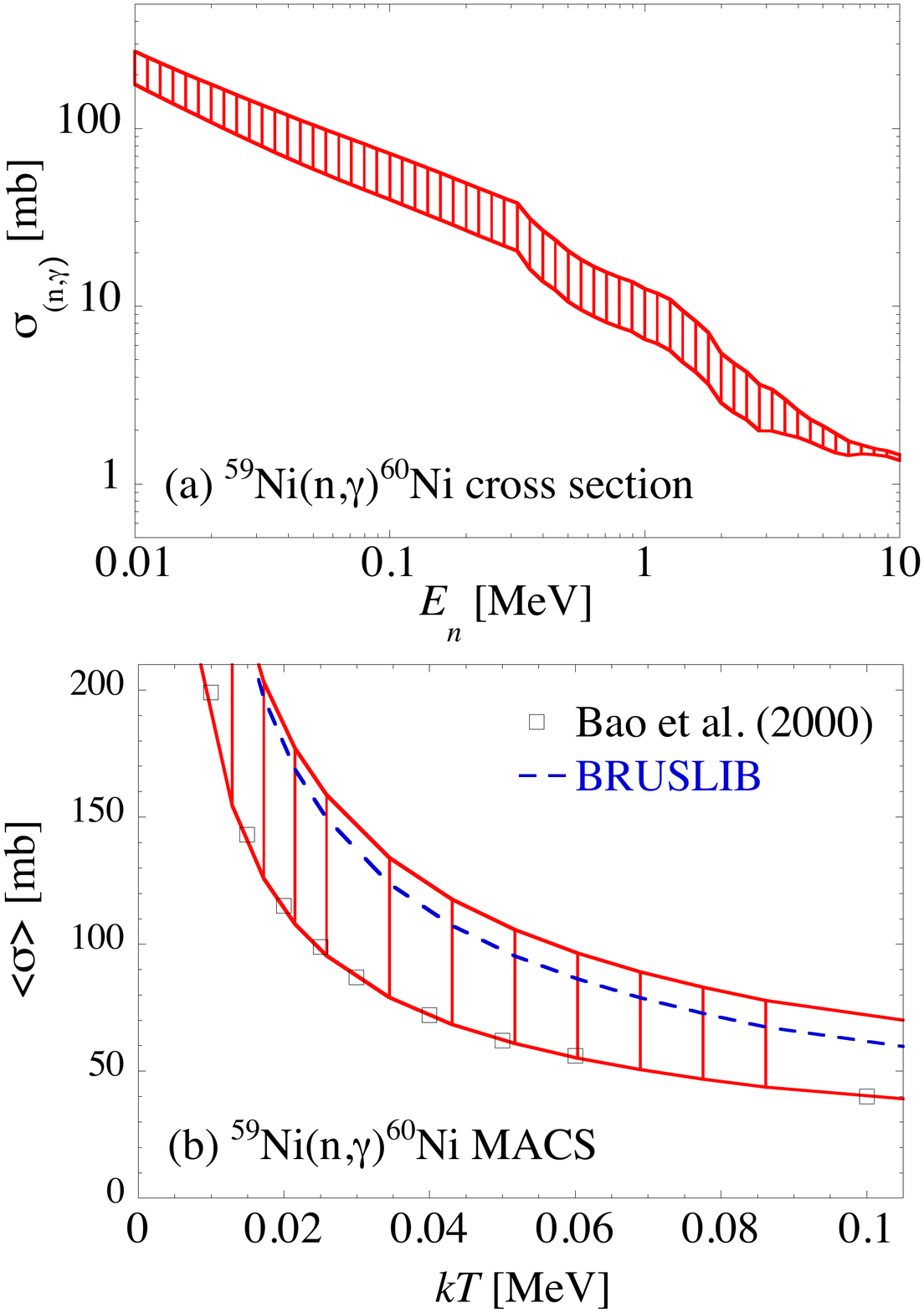}  
\caption{(Color online) (a) Calculated $^{59}$Ni$(n,\gamma)^{60}$Ni cross section. (b) same for the MACS (red solid lines). The open black squares correspond to the theoretical calculation recommended in Ref.~\cite{Bao00} and the blue dashed line to the Brussels library (BRUSLIB) prediction \cite{Bruslib}.}
\label{fig_Ni59ng}
\end{center}
\end{figure}

\subsection{Photon strength functions}
\label{sect_gsf}

The present experimental results have been analyzed here in light of the recent systematics of the $\gamma$SF obtained within the mean field plus QRPA calculations based on the finite-range Gogny D1M interaction \cite{Martini16,Goriely16b,Goriely18a}. When compared with experimental data and considered for practical applications, the mean field plus QRPA calculations need  some phenomenological corrections. These include a broadening of the QRPA strength  to take the neglected damping of  collective motions into account as well as a shift of the strength to lower energies due to the contribution beyond the 1 particle - 1 hole excitations and the interaction between the single-particle and low-lying collective phonon degrees of freedom (it should be noted that phonon coupling, in particular, may not be reduced to a simple broadening or shift of the $E1$ strength, as shown in Ref.~\cite{Achakovskiy15} for the Ni isotopes). Such phenomenological corrections  have been applied to the present Ni isotopes, as described in Ref.~\cite{Goriely18a} with an $E1$ damping width of 4.5~MeV and $M1$ damping width of 2~MeV. 
Such an $E1$ damping width is smaller than the systematics of $\Gamma_{E1}=7-A/45$~MeV deduced from photodata \cite{Goriely18a} due to the closed proton shell in Ni isotope leading to stiffer surface vibrations.
As a consequence, 
a factor of 2/3 on the overall $E1$ strength is required to reproduce the present peak photoneutron cross section in the GDR region. 

However, it remains unclear if the origin of this correction factor is specific to the closed shell Ni region or also found more systematically for light nuclei; more experimental photodata on light nuclei could help understanding this aspect.
A similar factor is applied to all the Ni $\gamma$SF. 

When considering the de-excitation, downward, strength function, $\overleftarrow{f_{X\lambda}}$, deviations from the photoabsorption strength can be expected, especially for $\gamma$-ray energies approaching the zero limit. In particular, shell model calculations predict a constant $E1$ strength at energies  typically below 5~MeV  \cite{Sieja17} as well as an increase of the $M1$ strength at decreasing energies approaching zero \cite{Schwengner13,Brown14,Sieja17,Schwengner17}. While the low-energy correction due to the $E1$ contribution has been shown to have a rather negligible impact on the $\gamma$SF and related radiative neutron capture cross section \cite{Goriely18a}, the impact of the low-energy enhancement of the $M1$ strength, the so-called $M1$ upbend, is far from being insignificant and strongly depends on its zero limit. The $\gamma$SF upbend is therefore of particular importance to constrain nucleosynthesis reactions. In the present paper, we will follow the same prescriptions as used in Ref.~\cite{Goriely18a}, {\it i.e.} the  final $E1$ and $M1$ strengths (hereafter denoted as D1M+QRPA+0lim) include the QRPA as well as the zero-limit contributions and are expressed as
\begin{eqnarray}
\overleftarrow{f_{E1}}(\varepsilon_\gamma) & = &  f_{E1}^{QRPA}(\varepsilon_\gamma) +f_0 U / [1+e^{(\varepsilon_\gamma-\varepsilon_0)}] \label{eq1}\\ 
\overleftarrow{f_{M1}}(\varepsilon_\gamma)  &=&  f_{M1}^{QRPA}(\varepsilon_\gamma) +  C~e^{-\eta \varepsilon_\gamma}
 \label{eq2}
\end{eqnarray}
where $f^{QRPA}_{X1}$ is the D1M+QRPA strength at the photon energy $\varepsilon_\gamma$, $U$ (in MeV) is the excitation energy of the initial de-exciting state, and $f_0=10^{-10}$~MeV$^{-4}$, $\varepsilon_0=3$~MeV, and $\eta=0.8$~MeV$^{-1}$ \cite{Goriely18a}. The value of $C\simeq 10^{-8}-10^{-7}~{\rm MeV^{-3}}$ is considered as a free parameter that is adjusted to the present data. An $M1$ zero limit $C=10^{-8}~{\rm MeV^{-3}}$ derived from shell-model calculations \cite{Goriely18a}, together with Eqs.~(\ref{eq1}-\ref{eq2}), was found to provide a rather good systematic description of available photoneutron data, average resonance capture data, Oslo $\gamma$SF as well as averaged radiative widths. Different shell-model interactions may provide however different predictions, and for medium-$A$ nuclei in the Fe region, larger values could be envisioned from previous Oslo measurements \cite{Crespo2016,Algin08}. For this reason, two different values are adopted in the present analysis, namely $C=3\cdot  10^{-8}$ and $10^{-7}~{\rm MeV^{-3}}$. The $M1$ zero limit  $C=10^{-7}~{\rm MeV^{-3}}$ is rather large with respect to shell-model predictions but is found in the present paper to be an upper limit that cannot be excluded in view of the new Ni experimental data at low energies, as shown below.

We compare in Fig.~\ref{fig_gsf} the Ni $\gamma$SF extracted from photoneutron and Oslo measurements with the D1M+QRPA+0lim dipole ($E1$+$M1$) $\gamma$-ray strength obtained with both values of $C$. It can be noticed that the predictions with $C=10^{-7}~{\rm MeV^{-3}}$ are compatible with the Oslo data at low energies and tend to be close to the lower limits, while the lower value of $C=3\cdot  10^{-8}~{\rm MeV^{-3}}$ in most cases underestimates the extracted $\gamma$SF at energies below 3-4~MeV, except for $^{64}$Ni. As far as the GDR region is concerned, the D1M+QRPA calculations are in relatively good agreement with the photoneutron data, even in the 10~MeV region where one can see extra $M1$ strength on top of the $E1$ component, as seen  in $^{61}$Ni.

\subsection{Radiative neutron capture cross sections}

On the basis of the theoretical $\gamma$SF described in Sect.~\ref{sect_gsf}, we now test the corresponding inputs in radiative neutron capture cross sections for nuclei for which cross sections have been measured. In addition to the $\gamma$SF, the radiative neutron capture is rather sensitive to the nuclear level densities. For this reason, five different nuclear level density models have been considered \cite{Koning08,Demetriou01,Goriely08,Hilaire12}, all of them being adjusted on experimental low-lying states as well as s-wave resonance spacings whenever available experimentally \cite{Capote09}.  
The theoretical $\gamma$SF as well as nuclear level densities used in TALYS also reproduce, within their uncertainties, the average radiative width from experiment or systematics, as discussed in Sect.~\ref{sect_gsf}.

TALYS predictions and experimental data are compared in Fig.~\ref{fig_ng} for the radiative neutron capture cross sections and in Fig.~\ref{fig_macs} for the Maxwellian-averaged cross sections (MACS). While the $\gamma$SFs with $C=3\cdot  10^{-8}~{\rm MeV^{-3}}$ tend to underestimate the low-energy Oslo data, they are seen  to reproduce rather well the neutron capture cross sections, except for $^{63}$Ni$(n,\gamma)^{64}$Ni which is clearly underestimated. Despite the fact that the D1M+QRPA+0lim strength  with $C=10^{-7}~{\rm MeV^{-3}}$ are rather lower limits of the measured $\gamma$SF (Fig.~\ref{fig_gsf}) below the neutron separation energy, they give rise to cross sections slightly larger than experimental values, except again for $^{63}$Ni$(n,\gamma)^{64}$Ni.  Similar conclusions can be drawn when looking at the comparison of MACS (Fig.~\ref{fig_macs}), although, in this case, the $\gamma$SFs with $C=10^{-7}~{\rm MeV^{-3}}$ clearly lead to MACS in agreement with data, except for $^{58}$Ni neutron capture.

Based on the above-described systematic study of the experimentally constrained $\gamma$SF (Eqs.~\ref{eq1}-\ref{eq2}), we can now apply the $\gamma$SF method to the still unknown estimate of the $^{59}$Ni$(n,\gamma)^{60}$Ni cross section. The corresponding cross section and MACS are shown in Fig.~\ref{fig_Ni59ng} where the shaded area corresponds to the uncertainties associated with the unknown value of the $M1$ upbend ({\it i.e.} the $C$ value) and the nuclear level density model. The theoretical MACS recommended in Ref.~\cite{Bao00} is seen in Fig.~\ref{fig_Ni59ng}(b) to correspond to our lower limit, while the Brussels library (BRUSLIB) prediction \cite{Bruslib} is closer to our upper limits. 

These comparisons show that experimental data constraining the $\gamma$SF at low energies, either through the Oslo method or photoneutron measurements, are of high importance for a relevant determination of the reaction cross section. Many uncertainties still affect the estimate of the low-energy tail of the GDR but the combination of relevant experimental data (like those derived in the present paper) with theoretical model (like the QRPA approach or the shell model) is little by little shedding light on the complex nuclear phenomena taking place during nuclear reactions and that remain of particular interest for astrophysical applications.

\section{Conclusions}
\label{sec-conclusions}
We presented results of the $(\gamma,n)$ cross section measurement for $^{58}$Ni, $^{60}$Ni, $^{61}$Ni, and $^{64}$Ni performed at the NewSUBARU synchrotron radiation facility. 
We have discussed radiative neutron capture cross sections and the MACS for Ni isotopes in terms of the D1M+QRPA dipole ($E1$+$M1$) $\gamma$-ray strength supplemented with the $M1$ upbend, which is constrained by the present $(\gamma,n)$ cross section and/or the Oslo data.    
With two choices of the zero-limit value for the $M1$ upbend ($C=3\cdot  10^{-8}{\rm MeV^{-3}}$ and $C=10^{-7}{\rm MeV^{-3}}$), consistencies of the model $\gamma$-ray strength function with the Oslo data for $^{59,60,64,65}$Ni, and known $(n,\gamma)$ cross sections and the MACS for 
$^{58,60,63,64}$Ni were discussed. 
In some cases, it remains difficult to reconcile $\gamma$SF and cross section data; it would be of interest to confirm the experimental average radiative width for such nuclei, but also to refine the systematics for the Ni isotopes for which no measurement is available.
Research along the present systematic study of the $\gamma$-ray strength function which describes the low-energy tail of the GDR with $(\gamma,n)$ cross sections, Oslo data, and others as experimental constraints helps understanding radiative neutron capture cross sections of direct relevance to the s-process nucleosynthesis, in particular, for radioactive nuclei at the s-process branching points. 

\section{Acknowledgments}
The authors are grateful to D. M. FIlipescu of the IFIN-HH and ELI-NP for providing $^{64}$Ni and $^{60}$Ni metal powders for the present experiment. They are also grateful to F. Kitatani and H. Harada of the JAEA and H. Ohgaki of the Institute of Advanced Energy, Kyoto University for making a $^{61}$Ni disk and a large volume LaBr$_3$(Ce) detector available for the experiment, respectively.  S.G. acknowledges the support from the F.R.S.-FNRS. 
H.U. acknowledges the support from the Premier Project of the Konan University. G.M.T.  and A.G. acknowledge funding from the Research Council of Norway, Project Grant Nos. 262952 and 263030, respectively.  A.C.L. acknowledges funding from ERC-STG-2014, grant agreement No. 637686.  D.S. acknowledges the support of Deutsche Forschungsgemeinschaft through grant No. SFB 1245.  This work was supported by the IAEA and performed within the IAEA CRP on ``Updating the Photonuclear data Library and generating a Reference Database for Photon Strength Functions'' (F41032).

\end{document}